\documentclass{ws-ijmpc}
\usepackage{graphics}

\newcommand{\diagon}{\texttt{\textsc{Diagon}}}
\begin{document}

\markboth{Jordan Kyriakidis}{A generic approach to electronic
  structure calculations in nanoscopic systems}

\catchline{}{}{}{}{}

\title{A GENERIC APPROACH TO ELECTRONIC STRUCTURE CALCULATIONS IN
  NANOSCOPIC SYSTEMS}

\author{JORDAN KYRIAKIDIS}

\address{
  Department of Physics and Atmospheric Science,\\
  Dalhousie University, Halifax, Nova Scotia, Canada, B3H 3J5\\
  http://soliton.phys.dal.ca}

\maketitle

\begin{history}
  \today
\end{history}

\begin{abstract}
  We outline a generic, flexible, modular, yet efficient framework to
  the computation of energies and states for general nanoscopic
  systems with a focus on semiconductor quantum dots.  The approach
  utilizes the configuration interaction method, in principal
  obtaining all many-body correlations in the system.  The approach
  exploits the powerful abstracting mechanisms of C++'s template
  facility to produce efficient yet general code.  The primary utility
  of the present approach is not in the resulting raw computational
  speed, but rather in minimizing the time from initial idea to final
  results.
\end{abstract}

\ccode{PACS Nos.: 07.05.Bx, 89.20.Ff, 73.21.La}

\section{Introduction}
The configuration interaction (CI) approach to electronic structure
calculations has the benefit of being conceptually simple and the
capability of being potentially exact---exact in the sense that all
eigenstates and energies of the given model system can be exactly
computed given sufficient computational resources.  The detriment of
the CI method is the great resources it requires.  Generally speaking,
state of the art calculations perform at 8--10 particles.  The reason
is that the CI method is essentially a direct diagonalization of the
system's Hamiltonian matrix, and the size of this matrix grows (at
least) exponentially with the number of particles.

Because of this, other techniques, such as quantum Monte Carlo (QMC)
and density functional theory, in all their various guises, are often
employed (and often required) in place of CI calculations.  Of these,
QMC and CI methods are the two primary tools in cases where
correlations are strong, where excited states in addition to ground
states are required, and especially where the many-body states
themselves (as opposed to the energies) are required.  QMC has great
potential for being far more efficient than CI, yet is also far more
complex both in concept and in practical aspects of coding.  Thus, QMC
may be viewed as being well suited for problems where a large number
of particles is required, or for stable production code which will be
applied to a specific fixed set of systems over a prolonged period of
time.

One may distinguish, however, between fast code written slowly---that
is, taking a long time to code, test, debug, and maintain---and slow
code written quickly.  QMC belongs in the former category, while CI
belongs in the latter.  Particularly for basic research, and perhaps
less so for applied research, a more useful interpretation of
``efficiency'' than the actual computation time is to include
development time as well.  Often, the most relevant time is that
between initial idea and final result, and by no means is this
duration necessarily dominated by raw computation time.  In this
sense, a properly structured CI base can be an important piece of a
research framework, particularly when flexibility is highly desired.

\section{The need for a flexible computing environment}
\label{sec:need-flex-comp}

Flexibility is particularly required for computations involving the
basic physics of synthetic nanostructures such as semiconductor
quantum dots.\cite{reiman02:elect.struc.quant.dots}  There are four
main reasons for this: the nonuniversality of synthetic structures,
the insufficiency of \textit{ab initio} methods for large systems, the
many different environmental couplings in heterogeneous solid-state
devices, and the highly tunable symmetries experimentally attainable.
In this section, we explore these issues by comparing them to atomic
systems, where very robust and highly efficient code is available.

In semiconductor quantum dots---for example, artificial
atoms\cite{kastner00:SET} defined by electrostatic (lateral)
confinement of electrons in a two-dimensional electron gas---each
manufactured device is unique and, for example, the confining
potential of two devices with ostensibly identical gate geometries and
identical numbers of confined particles, will nevertheless have unique
confining potentials, tunneling barriers, and other parameters
significantly affecting the transport, electronic, and spin properties
of the device.  Contrast this with atoms, where every distinct atomic
species is an indistinguishable particle; a large-scale, robust,
sophisticated, and efficient code base\cite{gonze02:_abinit}
eventually pays off since the resulting application is indefinitely
applicable to all future problems of the same atomic species.

In addition to the non-uniqueness of synthetic structures, many are
also large in size, and can contain, for example, several millions of
different atomic species arranged in heterogeneous geometries.  Not
only are current \textit{ab initio} methods insufficient to deal with
nonperiodic systems of such size, but often they are also undesirable;
numerical results on simpler effective Hamiltonians many times yield
more insightful, more intuitive results, particularly when studying
experimentally observed effects (decoherence, for example) whose
underlying physical mechanisms are unknown.  In such cases, numerical
results are more properly viewed as aids to theoretical analysis
rather than pure simulation.

In quantum nanoelectronic systems, and mesoscopic systems more
generally, the relevant length scale of the constituent
particles---for example, their de~Broglie wavelength, or their mean
free path---extends over the dimensions of the confinement potential.
In such cases, boundary effects or other couplings to various
environments often drive the observed physics.  These couplings
generally destroy quantum coherence and lead the system to more
classical-like behavior.  Experiments can now probe this boundary
between classical and quantum physics and the relevant physics can
often times be explored only numerically.  A numerical approach must
be flexible enough to investigate these quantum statistical systems,
neither periodic nor isolated, coupled to a heterogeneous and
fluctuating environment, and driven far from equilibrium.  Such a
purely quantum numerical framework does not currently exist.  Indeed
even the basic physical theory is often poorly understood.  In such
cases, particularly when only a few interacting electron are present
(but infinite bath degrees of freedom), flexibility usually (not
always) trumps raw processing power.

\section{The choice of programming language}
In terms of raw speed, simpler languages are usually superior.  This,
along with simple inertia, likely explains the dominance, or at least
the preponderance, of \textsc{fortran} for numerical code.  But,
generally speaking, simpler implies less expressive which in turn
implies less flexible.  If flexibility is to be put at a premium, then
a more expressive language is required.  This will generally come at a
cost of slower speed, but quicker development.

A practical compromise is a language where the programmer can decide
how much of an abstraction penalty to pay.  This
pay-only-for-what-you-use approach is well represented in the C++
programming language, and this is the language in which we have
developed the \diagon\cite{diagonNote} framework we describe below.

More important than efficiency, more important than flexibility,
correctness is the single most important requirement of all numerical
code.  An aid, but by no means a savior, to correctness is C++'s type
safety.  A constant can really be a constant without recourse to
preprocessor macros.  Pointers can point to (various) constant objects
or can be constant pointers themselves.  Parameters to functions can
be passed by value, by reference, or, most efficiently, by constant
reference.  These often obviate the need for raw pointers to memory
and all their error prone complexity.  Authors of a class have at
their disposal (and their discretion) a great deal of control over how
memory is precisely allocated, and how objects are created, assigned,
copied, or converted to other types.  These language features are
great tools in producing flexible yet robust numerical frameworks.

\subsection{Generic programming}
None of the above has anything to do with object-oriented
programming---the paradigm most frequently associated with C++.
Indeed C++ supports several programming styles, including procedural
and functional, as well as object-oriented.  However, in our opinion,
the single most useful feature of C++ in regards to numerical
computation is its generic programming facilities.  These can be seen
as a bridge between expressiveness on the one hand, and efficiency on
the other.  Particularly for scientific programming, recent
developments\cite{veldhuizen:_c++} in generic programming have clearly
shown that flexibility and abstraction need not necessarily incur a
run-time penalty and have moved the state of the art far beyond simple
parametrized types.

At its most basic level, generic programming separates data types from
algorithms.\cite{austern99:_gener_progr} Thus, for example, a single
sort algorithm can be written that can sort objects of arbitrary type,
including user-defined classes, so long as the expression $A < B$ is
defined for objects of those types.  And because of operator
overloading, the less-than operator can be defined by the author of
the data type.  Because of this, generic programming can be said to
make code forwards compatible in time, rather than simply backwards
compatible with previous versions.  If a specific type requires a more
efficient algorithm, that particular type can be made into a special
case through partial template specialization.

Importantly, this great flexibility can come with little or no
performance penalty---a crucial difference compared to dynamically
typed languages such as Python or Ruby.  The C++ generic facilities
produce statically typed compiled code.  In effect, a generic template
function is a program that writes programs and is essentially a
type-safe meta-programming facility.  Upon compilation, the compiler
takes a template function, and instantiates a specific version for
whatever data type is required.  Thus, a single sort template function
can produce several custom-made (by the compiler) versions in the
executable; one each for, say, integers, doubles, complex numbers, or
many-body state vectors sorted by energy.  The important fact is that
this happens at compile time, not run time, and so the produced code
can be very efficient.  The compiler can turn this single function
into an unbounded number of functions operating on types which the
original author could not have conceived.

The \diagon\ framework we describe below really only scratches the
surface of what generic programming can do.  For example, templates
can be used to keep track of dimensional
quantities,\cite{alexandrescu01:_modern_c++} so that, for example, the
system knows that length/time is a velocity and can give a
(compile-time) error when a length is assigned to a velocity, even
though all quantities are, say doubles.  In fact, scientific
programmers recognized quite early on\cite{barton94:_c++} the benefits
of generic programming.

As an example of the flexibility of generic programming coupled with
function overloading, we give here an example of their use in the
\diagon\ framework which we discuss more fully below.  Specifically,
we can consider the inner product of various state vectors describing
electrons in a quantum dot.  There are at least three types of states
that are generally required.  At the single-particle level, there are
the base single-particle orbital states $|\alpha\rangle$, where
$\alpha$ denotes a full set of quantum numbers, which we label
generically as \texttt{SPState}.  A many-body Fermion system is
described by antisymmetrised vectors $|\alpha_0, \alpha_1, \alpha_2,
\ldots\rangle$ (Slater determinants, labeled \texttt{AntiSymmState})
in accordance with the Pauli exclusion principle.  Thirdly, correlated
states $|\psi\rangle = \sum_i c_i |\alpha^i_0, \alpha^i_1, \alpha^i_2,
\ldots\rangle$ (induced by Coulomb interactions or spin
symmetry\cite{kyriakidis06:_fract_orbit_occup}) require a description
in terms of coherent superpositions of Slater determinants,
\texttt{LinCombState}.

The inner product of \texttt{SPState}'s clearly depends on the the
particular system under consideration.  However, the algorithmic
computation of inner products of either \texttt{LinCombState}'s or
\texttt{AntiSymmState}'s is essentially identical and independent of
the underlying \texttt{SPState}.  Thus, once supplied with an inner
product function for the particular \texttt{SPState}, the calculation
for inner products involving more complex state vectors can proceed
automatically, without a need for rewriting the functions.

The object-oriented solution to this problem is to define a class
hierarchy and to pass pointers (or references) to the functions.  At
run time, the appropriate inner product will be called.  Besides the
extra (run-time) cost involved in dereferencing objects for dynamic
polymorphism, this approach will almost certainly lead to an
ever-growing hierarchy of states, with commensurate costs in
maintenance and testing.  In addition, the core code base requires
altering with every new \texttt{SPState} introduced to the system; as
the number of states known to the system grows, maintenance, and
continued testing for correctness, becomes more and more of a burden.

In a generic approach, the code base remains small.  Here, one defines
generic containers \texttt{AntiSymmState<SPState>} and
\texttt{LinCombState<State>}\footnote{Note that a
  \texttt{LinCombState} can describe linear combinations of
  \texttt{AntiSymmState}'s with different sizes.  This would be
  required when particle number is not conserved, as, for example,
  when studying transport through a quantum
  dot.\protect\cite{vaz06:_dark_chann}} which can hold \emph{any} type
of state.  Provided the user implements a set of \texttt{SPState}'s
and defines single-particle inner-product functions, then inner
products involving many-body states need not be (re)written; the
compiler will write a custom version of the appropriate inner product
function.  The template functions themselves indicate the algorithm,
not the data types.

Because generic components can be combined as required, a small set of
generic components are capable of producing combinatorially many
functions.  Thus, with a small set of components, a large array of
composite objects can be defined, maximizing the flexibility of the
numerical approach.  This flexibility need not incur a run-time
penalty; all template instantiations are implemented at compile-time.
This is a tremendous advantage both in the speed of developing a
custom application from a set of generic components, as well as in
maintaining efficient code at run time.

We have implemented such a generic framework, \diagon, for computing
eigenstates and eigenvalues of semiconductor quantum dots for
arbitrary potentials, with arbitrary numbers of particles, and
arbitrary Hamiltonians.  This is a framework, meant for developing
applications rather than a tool for end users.  We describe the
framework below and show that it can be used to build flexible and
extensible applications without undue sacrifice on run-time
efficiency.

\section{The \diagon\ Framework}
The \diagon\cite{diagonNote} framework consists of a set of generic
components useful in the manipulation of many-body quantum states and
computations involving them.  There are currently components for
dealing with various types of Fermion states, components for computing
matrix elements of one and two-body operators, generators returning
proper spin eigenstates given a set of singly and doubly occupied
single-particle orbitals, and components for computing eigenstates and
eigenvalues of Hermitian operators.  These components are all generic
and the single-particle states themselves need to be provided.  We
give in this section an overview of the \diagon\ components.  In the
following section, we provide an example calculation for calculating
spectra and states of two-dimensional parabolic dots with spin-orbit
interactions.

\subsection{Generic quantum many-body states }
\label{sec:generic-quantum-many}

There are three classes of many-body states:
\texttt{AntiSymmState<SPState>} is the primary class for
antisymmetrised product states.  In real space, these are Slater
determinants.  Upon instantiation, the (generic) parameter
\texttt{SPState} must name an existing class encapsulating a known
single-particle state.  For example, in two dimensional
parabolically-confined quantum dots, the single particle states may be
the well-known Fock-Darwin\cite{jacak97:quant.dots} states
$|mns\rangle$.  Typically, the inner product between these two
single-particle states will also be provided.  In the Fock-Darwin
case, we simply have $\langle n'm's' | nms\rangle = \delta_{ss'}
\delta_{nn'} \delta_{mm'}$.  Operations are provided for creating and
destroying particles in these states, as well as computing inner
products if supplied with the inner product for the underlying
single-particle states.

In addition to the (user provided) single-particle states and the
(\diagon\ provided) \texttt{AntiSymmState}'s, a generic class
\texttt{LinCombState<State, Coeff>} is provided for encoding linear
superpositions of states, $|\mathtt{LinCombState}\rangle = \sum_\alpha
c_\alpha |\alpha\rangle$.  Here, the template parameter \texttt{State}
is the type of component states (which can be, for example,
\texttt{SPState}'s or \texttt{AntiSymmState}'s) and \texttt{Coeff} is
a template parameter encoding the type of the coefficients, which will
usually be real or complex numbers, but, since the class is generic,
could be of more exotic type.  Operations are provided for adding and
removing terms from the sum, for checking and setting normalization of
the state as a whole, for indexing a particular term, and for
iterating over all terms.  If class \texttt{State} defines an inner
product, then
$\langle\mathtt{LinCombState}|\mathtt{LinCombState}\rangle$ and
$\langle\mathtt{LinCombState}|\mathtt{State}\rangle$ are both defined.

The final class of many-body states is the \texttt{StateSet<State,
  Coeff>} class.  This class encapsulates a set of
\texttt{LinCombState<State, Coeff>}'s which all share a common basis.
It is used, for example, as a return type from a diagonalization
routine where each \texttt{LinCombState} is an eigenstate. It can also
be used to describe a set of spin eigenstates.  (See
Sec.~\ref{sec:generic-spin-states}.)  Operations are provided, for
example, to add or remove a basis vector to the \texttt{StateSet}, or
to add or remove a \texttt{LinCombState}.

With the above three generic components, one need only define a
particular single-particle state and an inner product, and many-body
states and linear superpositions of them are made available, and a
full suite of inner products and other manipulations and computations
are made available.

\subsection{Generic operator functions}
\label{sec:generic-operators}

A general quantum operator $\hat{\cal O}$ can be given a matrix form
with elements $\langle \psi_i | \hat{\cal O} | \psi_j\rangle$, with
$i, j$ running over all basis vectors which can be \texttt{SPState}'s,
\texttt{AntiSymmState}'s, or \texttt{LinCombState}'s.  The \diagon\
framework provides generic operators for both one and two body
operators.

A general one-body operator can be written in second quantized form as
\begin{equation}
  \label{eq:oneBodyOp}
  \hat{\cal O} = \sum_{p,q} {\cal O}_{pq} c^{\dag}_{p} c_q
\end{equation}
where $c^{\dag}_p$ creates a particle in state $|p\rangle$.  In the
\diagon\ framework, matrix elements of the operator $\hat{\cal O}$ are
implemented generically as
\begin{verbatim}
     oneBodyOp(bra, ket, matelem).
\end{verbatim}
Here, \texttt{bra} and \texttt{ket} can be any (combination) of the
three basic types of states in Sec.~\ref{sec:generic-quantum-many},
and \texttt{matelem} is a user-defined function evaluating ${\cal
  O}_{pq}$ in Eq.~(\ref{eq:oneBodyOp}).  That is, \texttt{matelem}
must have signature
\begin{verbatim}
     ReturnType matelem(SPState bra, SPState ket),
\end{verbatim}
where \texttt{ReturnType} is an arbitrary type.  The generic function
\texttt{oneBodyOp} returns whatever \texttt{matelem} returns.

We see that once the user defines single-particle properties, which
will be different from system to system, the \diagon\ framework
implements the many-body functionality, which is in a sense a
universal function of the single-particle physics.

Two-body operators are implemented generically in a similar way.  A
difference now is how spin is treated.  In particular, any
spin-independent two-body operator (\textit{e.g.}, the Coulomb
interaction) can be written as
\begin{equation}
  \label{eq:twoBodyOp}
  \hat{U} = \sum_{\stackrel{\scriptstyle i,j,k,l}{\sigma,\sigma'}}
  U_{ijkl} c^{\dag}_{i\sigma} c^{\dag}_{j\sigma'} c_{l\sigma'} c_{k\sigma},
\end{equation}
where the $U_{ijkl} = (ij|\hat{U}|kl)$ are the
coefficients\cite{negele98:_quant_many_partic_system} which must be
provided by the user, with all indices $i,j,k,l$ denoting
single-particle states.  This is again implemented as
\begin{verbatim}
     twoBodyOp(bra, ket, matelem)
\end{verbatim}
but the two states must now contain at least two particles
(\texttt{AntiSymmState}'s or \texttt{LinCombState}'s) and
\texttt{matelem} must have signature
\begin{verbatim}
     ReturnType matelem(SPState bra1, SPState bra2,
                        SPState ket1, SPState ket2).
\end{verbatim}
The generic function \texttt{twoBodyOp} returns whatever
\texttt{matelem} returns.

In Sec.~\ref{sec:exampl-parab-dots}, we describe a specific example
using the above generic components.  Before we do, however, we discuss
generic facilities \diagon\ provides for computing eigenstates of
total spin and other more general Hermitian operators.

\subsection{Generic spin states}
\label{sec:generic-spin-states}
Because the CI technique is so computationally intensive, it is
important to take advantage of every significant symmetry in the
system as this affords a possibility to block-diagonalize the
Hamiltonian matrix, drastically reducing the computational load.
Simple symmetries such as conservation of spin or angular-momentum
projection along a given axis ($S^z_{\mathrm{tot}}$ or
$L^z_{\mathrm{tot}}$, say) are simple to implement since these
symmetries do not produce correlations and their conservation can
always be encoded in a single Slater determinant.  Other
symmetries---total spin, $S^2_{\mathrm{tot}}$, being the most
prominent---\emph{do} induce correlations, and a single Slater
determinant (\texttt{AntiSymmState}) cannot in general be written down
in which $S^2_{\mathrm{tot}}$ is a good quantum number.

In such cases, a correlated basis may be used which preserves the
many-body symmetry.  This, in general, would require a
prediagonalization step.\cite{wensauer04:config.inter.method.fock}
However, the SU(2) symmetry of spin, along with its higher-dimensional
representations, allows all eigenstates of spin to be written down for
arbitrary orbital configuration, essentially relying on the
appropriate products of Clebsch-Gordon
coefficients.\cite{helgaker00:_molec_elect,rontani06:_full_config_inter}

Such a facility is provided in \diagon\ through the \texttt{spinGen}
generic function,
\begin{verbatim}
     spinGen(AntiSymmState<SPState> config, int twoS, int twoSz),
\end{verbatim}
which returns a \texttt{StateSet}.  (See
Sec.~\ref{sec:generic-quantum-many}.)  Each element of the
\texttt{StateSet} is a \texttt{LinCombState<AntiSymmState<SPState>>}
which is an eigenstate of $S^2_{\mathrm{tot}}$ with the appropriate
spin quantum number.  Input to \texttt{spinGen} is two times the spin
$S$ and two times the projection $S_z$.  (This is to keep the inputs
integers.)  Input is also the orbital configuration as an
\texttt{AntiSymmState<SPState>}.

\subsection{Generic diagonalization}
\label{sec:gener-diag}
The CI method eventually requires a diagonalization.  Currently,
\diagon\ employs the uBLAS linear algebra library of the
Boost\footnote{Boost (http://www.boost.org) provides a collection of
  free peer-reviewed C++ libraries with an emphasis on generics and
  portability.} project, along with a bindings library allowing C++ to
directly interface with the LAPACK algorithms.  This can be extended
to other diagonalization routines without much trouble.  The function
has signature
\begin{verbatim}
     vector<double> diagon(Matrix H,
                           StateSet<State, Coeff> eigenVecs,
                           size_t numEigs),
\end{verbatim}
where \texttt{H} is a uBLAS matrix.  The final two arguments are
optional.  If the first is provided, then the eigenvectors of
\texttt{H} are calculated and placed in \texttt{eigenVecs}.
Otherwise, only eigenvalues are computed.  The final argument
\texttt{numEigs} indicates how many eigenvectors and eigenvalues to
compute.  If omitted, all are computed.  The function itself returns
the eigenvalues of \texttt{H} in a \texttt{vector<double>}.

To construct \texttt{H}, one would normally call \texttt{oneBodyOp}
and/or \texttt{twoBodyOp} for each of the elements.  To aid in this, a
function \texttt{matrixOp} is provided
\begin{verbatim}
     Matrix matrixOp(basis, matelem)
\end{verbatim}
which returns the uBLAS matrix obtained by applying \texttt{matelem}
(Sec.~\ref{sec:generic-operators}) to each of the basis vectors in
\texttt{basis}.  This function calls \texttt{oneBodyOp} and
\texttt{twoBodyOp} as appropriate.

As mentioned above, the primary purpose of the \diagon\ framework is
to provide flexible, generic tools to aid development.  This aspect
was placed at a higher premium than pure computational efficiency,
although the generic nature of the framework is very well suited to
producing efficient code as well.  In the following penultimate
section, we provide an example illuminating the strengths of the
\diagon\ framework.

\section{Example: Parabolic dots with spin-orbit interactions}
\label{sec:exampl-parab-dots}
Using the generic framework \diagon, a complete diagonalization
program can be set up and run remarkably quickly.  Once the basic
(non-generic) components are provided, the package, at compile-time,
produces a custom-made set of classes and functions dealing with
linear superpositions of many-body states.  These can be used as a
simple diagonalization to obtain spectra, or the eigenstates can be
used for further computations in, for example, problems of quantum
dynamics and decoherence, where the actual states are required, and
where correlations in the system play an important role.  In this
section, we outline a diagonalization procedure to illustrate the use
of the \diagon\ framework.

We look specifically with at a two-dimensional quantum dot in a
GaAs/AlGaAs heterojunction parabolically confined in the plane and in
the presence of both spin-orbit interactions and a magnetic field
perpendicular to the plane of the dot.\cite{jacak97:quant.dots} The
Hamiltonian may be written as
\begin{equation}
  \label{eq:FockDarwinHamil}
  \hat H = \hat H_{\mathrm{qd}} + \hat H_{\mathrm{so}},
\end{equation}
where the quantum dot Hamiltonian is given by two harmonic oscillators
plus a Zeeman term
\begin{equation}
  \label{eq:qdotH}
  \hat H_{\mathrm{qd}} = \hbar\Omega_+ \left(a^\dag a +
    \frac{1}{2}\right) + \hbar\Omega_- \left(b^\dag b +
    \frac{1}{2}\right) + g \mu_B B^z S^z,
\end{equation}
with $\Omega_{\pm} = [(\omega_c^2 + 4 \omega_0^2)^{1/2} \pm 1] / 2$.
Here, $\omega_0$ is the confinement frequency characterizing the
parabolic confinement,\cite{jacak97:quant.dots} and $\omega_c = e B^z
/ (m c)$ is the cyclotron frequency.  The final term in
Eq.~(\ref{eq:FockDarwinHamil}) is the spin-orbit interaction.  We take
a linearized model including both Dresselhaus and Rashba
terms,\cite{Zawadzki04:Spin-splitting} respectively given by $\hat
H_{\mathrm{so}} = \beta (-\sigma_x p_x + \sigma_y p_y) + \alpha
(\sigma_y p_x - \sigma_x p_y)$, where $\sigma_k$ are the Pauli
matrices, and $\beta$ and $\alpha$ are respectively the Dresselhaus
and Rashba coefficients.  In terms of the Bose operators of
Eq.~(\ref{eq:qdotH}), we can write
\begin{equation}
  \label{eq:Hso}
  \hat H_{\mathrm{so}} = \Lambda S_+ \left[ \Omega_- (\alpha b^{\dag} +
    i \beta b) - \Omega_+ (\alpha a + i \beta a^{\dag}) \right] +
  \mathrm{h.c.},
\end{equation}
where $\mathrm{h.c.}$ is the Hermitian conjugate, $S_+$ is the spin
raising operator, and $\Lambda^2 = (\hbar m / 2) / (\omega_c^2 + 4
  \omega_0^2)^{1/2}$.

The objective in this example is to diagonalize
Eq.~(\ref{eq:FockDarwinHamil}) in the basis of the eigenstates of
Eq.~(\ref{eq:qdotH}), given by the well-known\cite{jacak97:quant.dots}
Fock-Darwin states $|nms\rangle$.

\begin{figure}
  \centering
  \resizebox{4in}{!}{\includegraphics{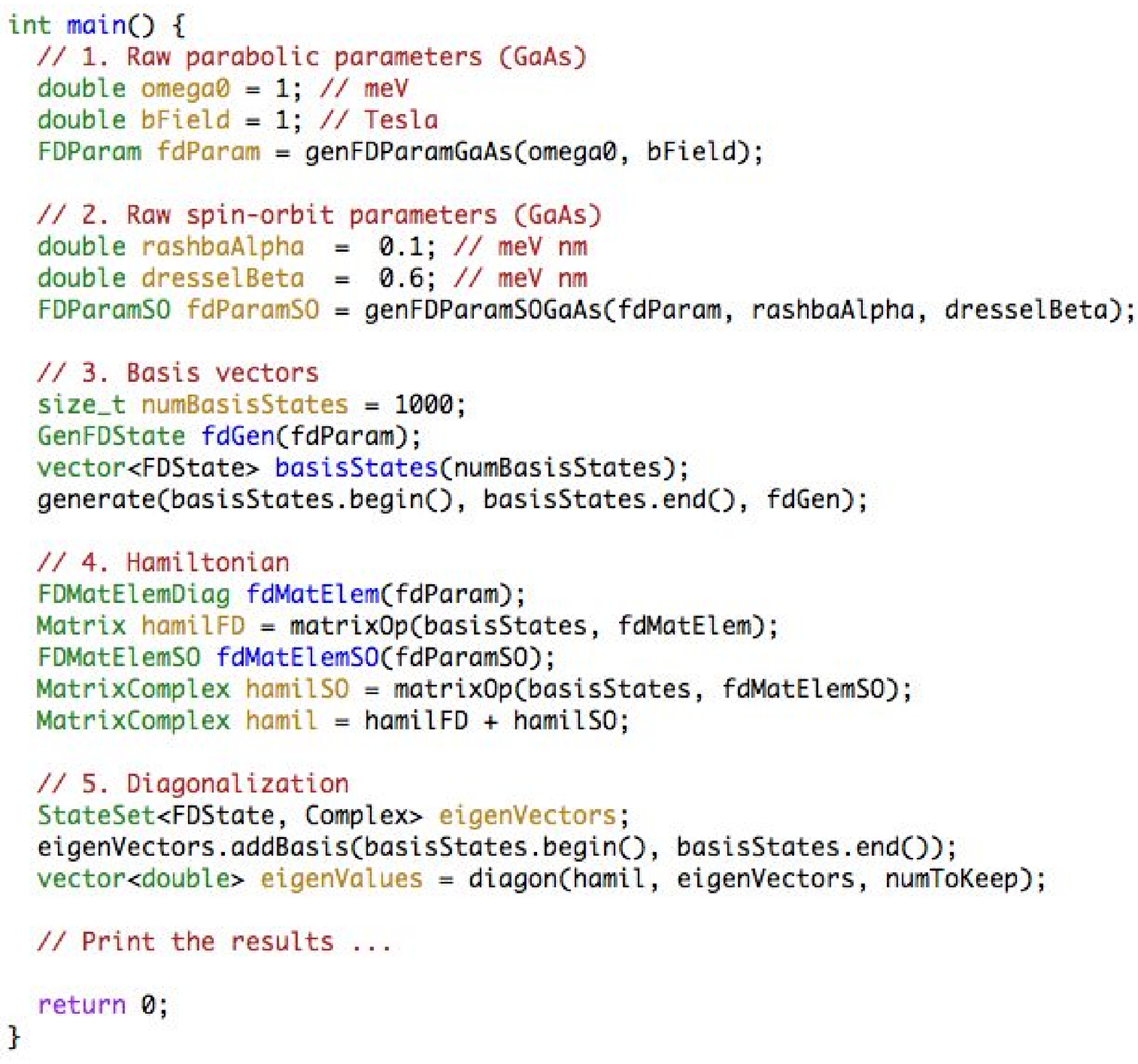}}
  \caption{Example code\protect\cite{diagonNote} showing minimal
    \protect\texttt{main()} function for computing spin-orbit
    eigenvalues and eigenvectors.  There are only 21 lines of code.}
  \label{fig:example-code}
\end{figure}

Figure~\ref{fig:example-code} shows a minimal function which does this
using the \diagon\ framework.  The example is meant only for
illustrative purposes; header files and additional comments have been
stripped for brevity.  Section~1 of the code simply creates parameters
for Eq.~(\ref{eq:qdotH}); the \texttt{genFDParamGaAs()} function takes
$\omega_0$ and the external field and computes $\Omega_{\pm}$ and the
Zeeman energy in Eq.~(\ref{eq:qdotH}) using GaAs material parameters.
Section~2 performs a similar task for Eq.~(\ref{eq:Hso}).

Section~3 generates the basis states with which to perform the
diagonalization.  The object \texttt{fdGen()} is a function which is
instantiated with the \texttt{fdParam} object; successive calls of
\texttt{fdGen()} return the Fock-Darwin state $|nms\rangle$ with the
next-highest energy.  Thus calling \texttt{fdGen()} 1000 times will
yield the 1000 lowest-energy states with the given material and model
parameters.  The line
\begin{verbatim}
     generate(basisStates.begin(), basisStates.end(), fdGen);
\end{verbatim}
does exactly this.  The basis states are then stored in the vector
\texttt{basisStates}.

Section~4 of Fig.~\ref{fig:example-code} creates the Hamiltonian
matrix, Eq.~(\ref{eq:FockDarwinHamil}), using the \texttt{matrixOp}
function of Sec.~\ref{sec:gener-diag}.  The function objects
\texttt{fdMatElem} and \texttt{fdMatElemSO} are each to be called with
two arguments (Fock-Darwin states, \texttt{FDState}) and return matrix
elements of $\hat H_{\mathrm{qd}}$ and $\hat H_{\mathrm{so}}$
respectively.  Thus, these are the \texttt{matelem} objects described
in Sec.~\ref{sec:generic-operators}.

Finally, in section 5 of Fig.~\ref{fig:example-code}, the
diagonalization occurs.  A \texttt{StateSet}
(Sec.~\ref{sec:generic-quantum-many}) is first created and the basis
states are added to it.  Then, \texttt{diagon}
(Sec.~\ref{sec:gener-diag}) is called and the eigenvalues are placed
in the vector \texttt{eigenValues}, whereas all the eigenstates, each
an orthogonal superposition of the basis vectors, are placed in the
\texttt{StateSet} \texttt{eigenVectors}.  These results can then be
printed, or otherwise processed.

We stress that this simple example is meant for illustrative purposes
only.  Many of the functions in Fig.~\ref{fig:example-code} take
optional arguments and support different interfaces and much of the
\diagon\ framework has not been explicitly mentioned in this example.
Nevertheless, it does illustrate how a properly constructed generic
framework can support a flexible computing environment.  For example,
elliptic dots could easily be added to the above example.  One would
simply need to define an additional \texttt{matelem} function object
which computes the appropriate matrix elements, and an additional
parameter object containing the eccentricities and so forth.  This
additional term could then be added to the Hamiltonian through the
\texttt{matrixOp} function.  Additionally adding Coulomb interactions
among the particles would proceed along essentially identical lines.

\section{Conclusions}
We have outlined an approach to the CI method which utilizes the
generic programming facilities provided by the C++ programming
language.  The general idea is that much of the CI machinery is
independent of the actual single-particle states used.  A generic
approach allows one explicitly separate algorithms and data types and
allows a great deal of code reuse.  This has been implemented in the
\diagon\ framework, which focuses on (but is by no means restricted to)
two-dimensional semiconductor quantum dots.  A generic approach such
as this offers a good compromise between rapid development and
flexibility on the one hand, and efficient code on the other.

\section*{Acknowledgments}
\label{sec:acknowledgments}
This work is supported by NSERC of Canada and by the Canadian
Foundation for Innovation.


\end{document}